\documentclass[5p,twocolumn]{elsarticle}
\usepackage[version=3]{mhchem}
\usepackage{hyperref}
\usepackage{lineno}
\usepackage{textcomp}
\usepackage{amssymb}

\journal{Nuclear Inst. and Methods in Physics Research, B}

\begin{document}

\begin{frontmatter}

\title{Understanding radioactive ion beam production at	ISAC through yield measurements and simulations}

\author[addresslabel1,addresslabel2]{Peter Kunz\corref{cor1}}
\ead{pkunz@triumf.ca}
\cortext[cor1]{Corresponding author. Tel.: +1 (604) 222-7690}

\author[addresslabel1,addresslabel3]{Jens Lassen}
\author[addresslabel2]{Corina Andreoiu}
\author[addresslabel4]{Fatima H. Garcia}
\author[addresslabel1,addresslabel2]{Hua Yang}
\author[addresslabel1,addresslabel5]{Valery Radchenko}

\address[addresslabel1]{TRIUMF, 4004 Wesbrook Mall, Vancouver, BC V6T 2A3, Canada}
\address[addresslabel2]{Department of Chemistry, Simon Fraser University, Burnaby, BC V5A 1S6, Canada}
\address[addresslabel3]{Department of Physics, Simon Fraser University, Burnaby, BC V5A 1S6, Canada}
\address[addresslabel4]{Lawrence Berkeley National Laboratory, Nuclear Physics Division, Berkeley, CA, USA}
\address[addresslabel5]{Department of Chemistry, University of British Columbia, 2036 Main Mall, Vancouver, BC, V6T 1Z1, Canada}

\begin{abstract}
The high-intensity proton beam of the TRIUMF 500 MeV cyclotron offers unique opportunities to produce rare isotopes by irradiating a variety of targets. In particular, the ISAC (Isotope Separation and ACceleration) facility provides the infrastructure to deliver customized rare ion beams for fundamental research in nuclear physics, astrophysics, material science, and nuclear medicine.
A continuous effort is made to develop new radioactive ion beams (RIB) and improve their intensity or purity, properties that depend strongly on the type of target, operating conditions, the ion source, and beam transport efficiency.
Yield data and theoretical production rates based on FLUKA and G\textsc{eant}4 simulations are collected in the \textit{TRIUMF Isotope Database}, providing a valuable resource for RIB development and experiment planning.
This paper introduces the newly upgraded \textit{TRIUMF Isotope Database} and presents examples on how it's resources can be used to plan experiments and understand the origin of certain RIB.
In particular we discuss the collection of $^{155}$Tb which is an important component of medical isotope research at TRIUMF.

\end{abstract}


\end{frontmatter}


\section{Introduction}

The Isotope Separation and ACceleration (ISAC) facility at TRIUMF \cite{dilling_isac_2013} offers a wide range of rare isotope beams for research in a variety fields from basic nuclear research to life science applications. Radioactive isotopes are generated mainly through spallation, fragmentation and fission reactions in suitable target materials with protons (p+) from the TRIUMF cyclotron.

The ISAC facility is based on the ISOL (Isotope Separation OnLine) technique \cite{al-khalili_isotope_2006}. Radioactive ion beams are extracted from \textit{thick} targets composed of a variety of refractory materials.
The production rate of specific isotopes $(Z,A)$ is determined by their production cross section $\sigma$ for the given p+ beam energy $E$, the p+ beam intensity $\Phi_{p+}$ and the target thickness $N_M$ where $M$ indicates the target material.
\begin{equation}
	\label{eq:prod}
	P(Z,A,E,M) = \Phi_{p+}(E)\; \sigma(Z,A,E,M)\;  N_M
\end{equation}
The p+ beam energy at ISAC is fixed at 480~MeV, but production rates can be optimized through the choice of target material and thickness.
The p+ beam intensity is limited to 100~$\mu$A though the permissible beam intensity is constrained by ability of the target dissipate the deposited beam power without exceeding the nominal target operating temperature.
For example, a typical high-power tantalum metal foil target \cite{dombsky_isac_2003} can accept up to 70~$\mu$A p+ at a thickness of 0.14~mol/cm$^{2}$ and an operating temperature of 2300~\textdegree C. A composite ceramic uranium carbide target \cite{kunz_composite_2013} with a typical thickness of 0.05~mol uranium/cm$^{2}$ can take up to 20~$\mu$A p+ at 1950~\textdegree C. The operating temperature is determined by the vapour pressure of the target material.
The pressure inside the target needs to be kept at high-vacuum levels ($< 10^{-5}$mbar) to preserve the ion source efficiency and structural integrity over run periods of up to 4 weeks.

The release of reaction products from the target is governed by their diffusion and effusion properties. The speed of both processes depends exponentially on target-specific diffusion constants and adsorption enthalpies respectively as well as the temperature.
Therefore, the volatility and release efficiency of isotopes through diffusion and effusion generally benefits from high operating temperatures, in particular with regard to short-lived isotopes. 

Combined, all the parameters discussed above contribute to the ion beam yield, in units of isotopes or ions per second, as defined in Eq.~\ref{eq:yield}. It is the product of the production rate (Eq.~\ref{eq:prod}) times efficiency factors for release $\epsilon_R$, ionization $\epsilon_I$, and beam transport $\epsilon_T$. 
\begin{equation}
	\label{eq:yield}
	Yield(Z,A) = P(Z,A)\cdot \epsilon_R\cdot \epsilon_I\cdot \epsilon_T
\end{equation}

Several methods are available to determine the yield experimentally. Ion beam currents higher than 1~pA can be measured directly with Faraday cups which are part of the diagnostic capabilities of the ISAC electrostatic beamline network \cite{bricault_rare_2013}. 
A Multiple-Reflection Time-Of-Flight Mass Separator (MR-TOF-MS) with high resolving power has been added to the experimental infrastructure at ISAC. 
It can separate isobaric beam components and identify low-intensity beams with rates of less than 0.1~isotopes/s \cite{reiter_improved_2019}. RIB of more than 1~isotope/s are usually characterized with the \textit{ISAC Yield Station} which is capable of identifying and quantifying multiple radioactive beam components simultaneously using $\alpha$, $\beta$ and high-resolution $\gamma$ spectroscopy \cite{kunz_nuclear_2014}.

The results provide important information for planning and evaluating the feasibility of experiments at the ISAC facility. After thorough evaluation, they are made accessible through the \textit{TRIUMF Isotope Database} \cite{isacyielddb}. 
Specific features of the online database are outlined in section \ref{idb}. 
The database is not just a valuable resource for experiment planning but also a research tool that provides new insights for radioactive ion beam development by correlating a wide range of experimental and theoretical data.

\section{Isotope Database}
\label{idb}

\begin{figure}
 	\resizebox{1.0\columnwidth}{!}{%
 		\includegraphics{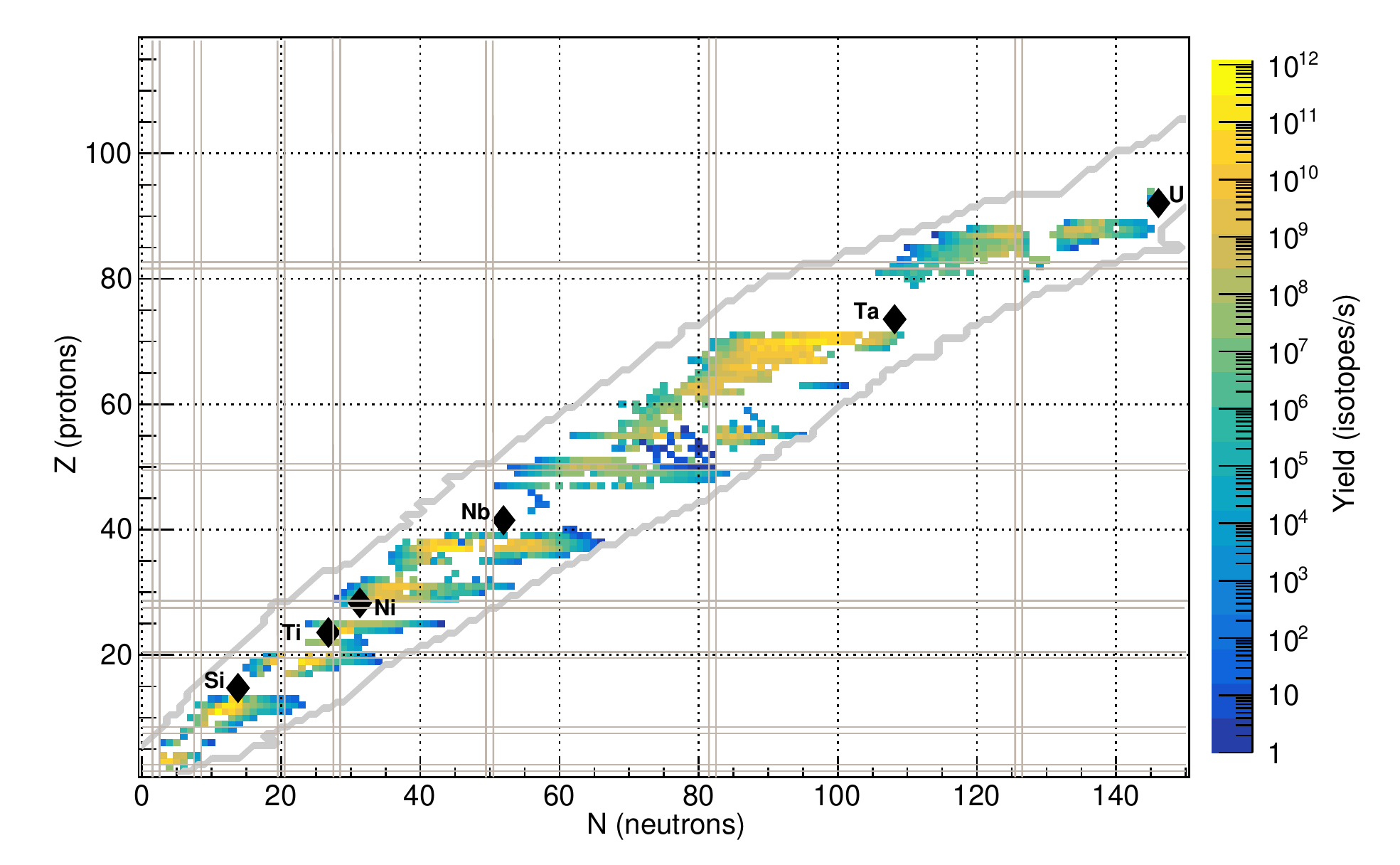}
 	}
 	\caption{The chart summarizes the most intense beams of close to 1000 radioactive isotopes extracted from ISAC targets. It was generated with a script that queries the \textit{TRIUMF Isotope Database} for the highest yield of each recorded isotope from a wide variety of targets. The most common target materials are indicated by {\large$\blacklozenge$}.}
 	\label{fig:allyields}       
 \end{figure}

The \href{https://yield.targets.triumf.ca}{\textit{TRIUMF Isotope Database}} \cite{isacyielddb} is maintained as a publicly accessible website based on open source platforms Node.js\textsuperscript{\textcopyright} serving as user interface and the SQL-based database server MariaDB for data management.
The web interface provides options for users to query and extract public information on RIB yields, production rates, target configurations and operating conditions. 
The database also contains not publicly available legacy and development data that can be accessed through custom tools to extract and process data sets, for example, to generate summaries, as shown in Fig.~\ref{fig:allyields}, extract systematic trends from development results or perform a comparative analysis of different theoretical production rate models and experimental results.

\subsection{Yield Data}
\label{yield_data}

Yield measurements are performed prior to each scheduled RIB experiment to optimize beam delivery and to establish whether the experiment is feasible with the given RIB intensity and composition.
Also, developing new RIB and the improving established beams requires systematic measurements. 
These results are also added to the database but usually for internal use only.

A yield data search can be refined though the following parameters: proton number, mass range, isomeric state, charge state, target ID, target material, ion source and measurement date. 
A database query can be customized by freely combining any of those parameters.
For example, if an experimenter would be interested in a specific isotope with A=155, a simple search by setting the mass number input to 155 generates a list of isotopes defined by mass over charge ratio (A/q), target material and ion source. 
For each row a yield based on the average of plausible results\footnote{Only results obtained under regular operating conditions are included. The database may contain entries for unusual operating conditions or development data. Those are usually not flagged for averaging or public viewing.} is given. 
The list may not only contain entries on ion beams with A/q=155 but also charge states higher than 1+ or molecular beams. 
There are options export the search result or display details on all or single entries. 
The links to details provide information on individual measurements including date, charge state, proton beam current, ion source and optional comments. 
Furthermore, a link to a description of the specific target that was in operation during the measurement is provided. This information can also be accessed through the "Target Search" option.

\subsection{Simulation Data}
\begin{figure}
	\resizebox{0.9\columnwidth}{!}{%
		\includegraphics{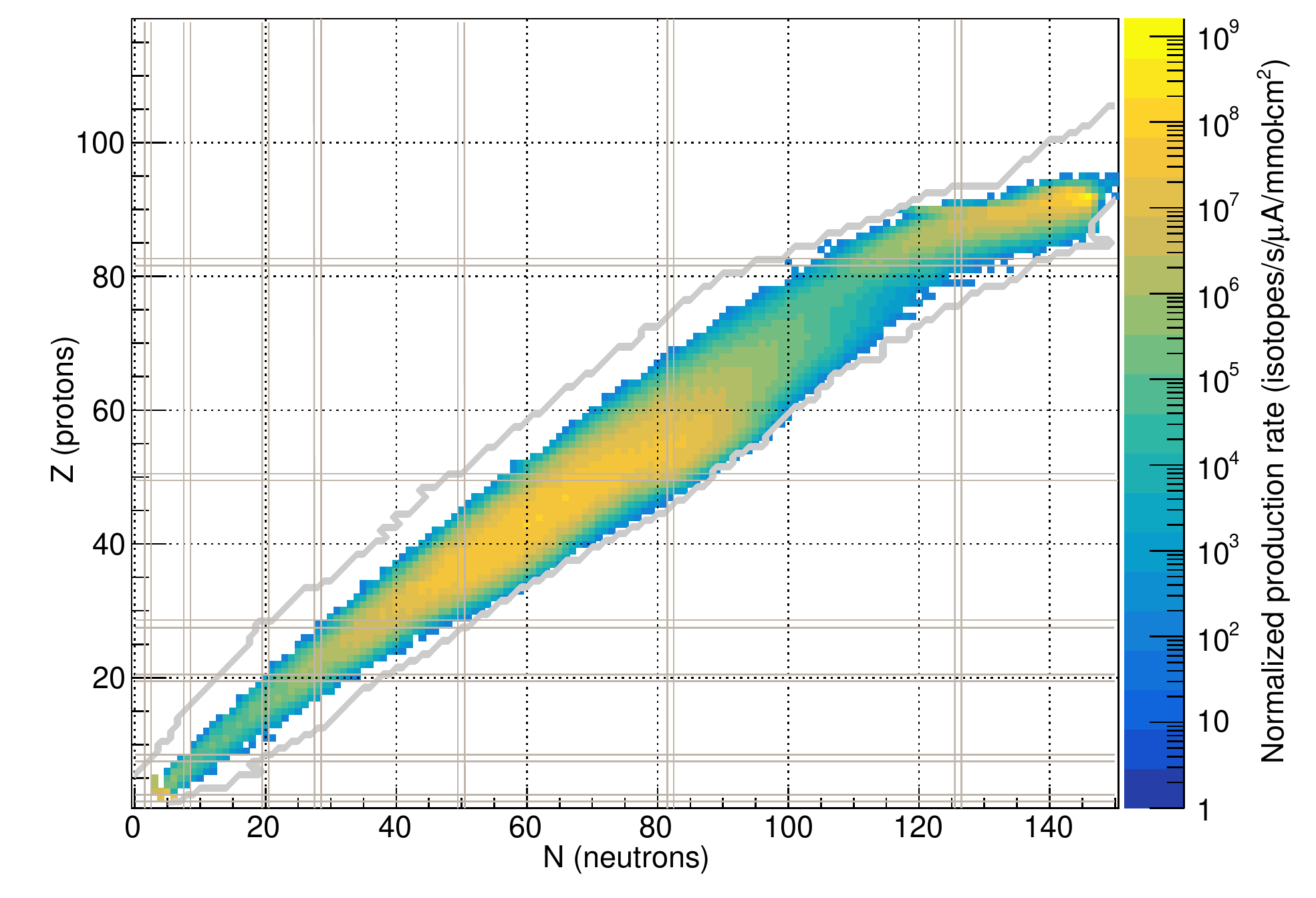}
	}
	\caption{G\textsc{eant}4 10.06 simulation of isotope production rates from a typical ISAC uranium target.}
	\label{fig:g4natu}       
\end{figure}

The option "Simulation Data Search" allows to look up data on calculated production rates for the most common target materials using the Monte Carlo simulation packages FLUKA \cite{bohlen2014211} and  G\textsc{eant}4 \cite{allison_recent_2016} with the Liege Intra-Nuclear Cascade Model \cite{rodriguez_2017}.
The simulations model a simplified target geometry, a cylinder with a diameter of 19~mm and a length of 500~mm with a standardized target thickness of 0.05~mol/cm$^2$. It is set by filling the cylinder with target material of matching density. 
These parameters, dimensions as well as target thickness, are fairly close to typical uranium targets. To be consistent the same boundary conditions have been applied to all simulations.
The typical thickness of lighter target elements such as Ta and Si can be about 3 to, respectively, 10 times higher. Beam power deposition and proton scattering are comparable for all targets so that scaling of production rates remains reasonably accurate.
The database contains normalized production rates in units of $isotopes/s/\mu A/mmol\cdot cm^2$ for isotopes produced from the most common ISAC targets with 480~MeV protons. The rates can be scaled to the expected target thickness ($mmol/cm^2$) and proton beam current. As an example, the data set of a  G\textsc{eant}4 simulation using the QGSP-INCLXX-HP physics list in combination with the ABLA31 evaporation code is displayed in Fig.~\ref{fig:g4natu}.

As with "Yield Search", the "Simulation Data Search" can be refined by combining a variety of parameters to obtain user-specific data sets. 
For example, search results can be used to compare specific production rates from different simulation models with yields, find peak production rates or correlate production rate trends with established yields. 
The information can help to determine the best target, the feasibility and scope of experiments or potential intensities of isobaric RIB components.
How the \textit{TRIUMF Isotope Database} is employed to plan an ISAC experiment is described in section \ref{ex} on a specific example.

\section{Collection of medical isotope $^{155}$Tb}
\label{ex}

For nuclear medicine applications $^{155}$Tb is collected at the \textit{ISAC Implantation Station} by implanting an RIB at A/q=155 in a custom vacuum chamber . 
Procedure and applications are described in detail in references \cite{kunz_medical_2020} and \cite{fiaccabrino2021}.
A crucial part of planning the experiment is to determine the most suitable target and the beamtime required to collect the desired amount of $^{155}$Tb.
A "Simulation Data Search" reveals that tantalum targets provide much higher production rates for neutron-deficient lanthanides than the alternative uranium. 
A query for isotopes at A=155 from Ta targets indicates from FLUKA and  G\textsc{eant}4 data sets that the peak production at this mass is around Z=68, $^{155}$Er. The ISAC mass separator transmits all isobars at A/q=155 with equal efficiency. 
As shown in Eq.~\ref{eq:decay_tb155}, radioactive isotopes produced at the highest rates decay mainly through $\beta^{+}$ / electron capture and eventually built up activity of the relatively long-lived $^{155}$Tb (T$_{1/2}$ = 5.32 d).
\begin{equation}
	\label{eq:decay_tb155}
	\begin{split}
		\ce{^{155}Er ->[{\beta^{+}/ec}][{5.3 m}] ^{155}Ho ->[{\beta^{+}/ec}][{48 m}] ^{155}Dy ->[{\beta^{+}/ec}][{9.9 h}] ^{155}Tb}
	\end{split}
\end{equation}
Therefore, the most efficient way to produce high activities of $^{155}$Tb in the shortest possible beamtime is to collect the isobaric RIB components Er, Ho and Dy which are produced at higher rates than $^{155}$Tb.

\begin{figure}
	\resizebox{1.0\columnwidth}{!}{%
		\includegraphics{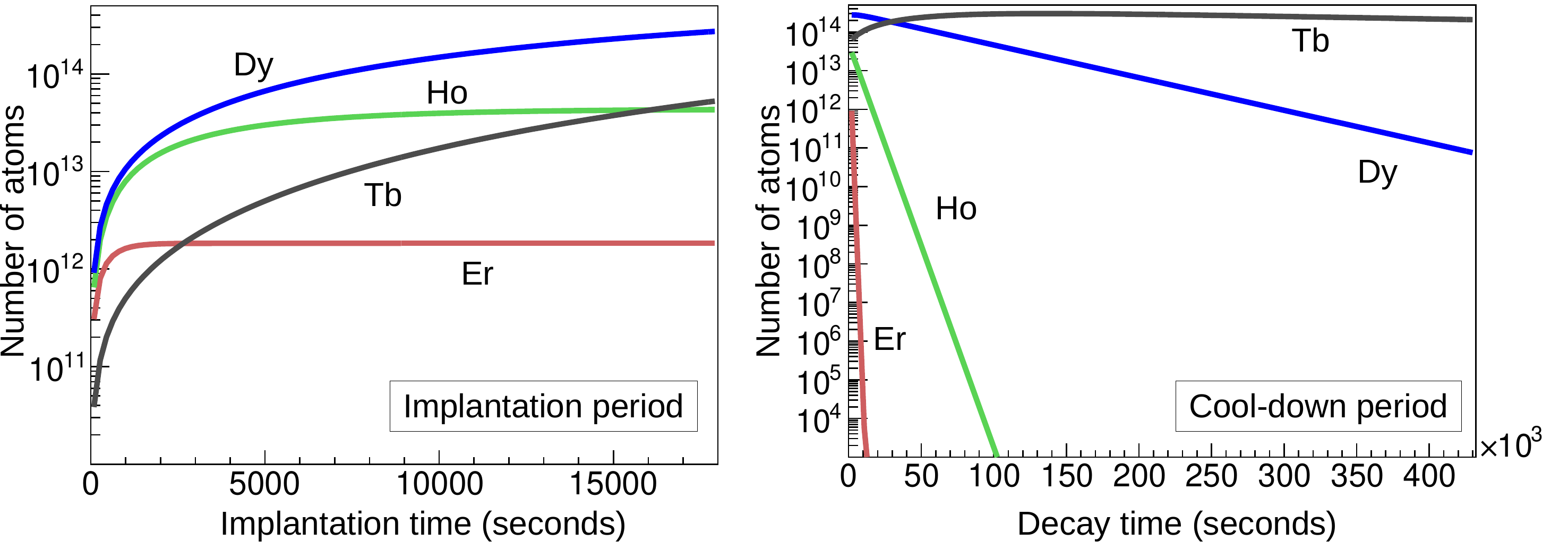}
	}
	\caption{Left: Build-up of isotopes based on measured yields of significant A/q = 155 ion beam components over scheduled implantation period. Right: Decay of accumulated activity over a 5-day cool down period.}
	\label{fig:tb155}       
\end{figure}

The $^{155}$Tb collection process consists of two stages shown in Fig.~\ref{fig:tb155}. During the implantation period all isobaric RIB components at A/q=155 are implanted in the collection vessel, including the most intense beams of Er, Ho and Dy. 
The implantation time is approximated for a specific $^{155}$Tb activity based on the individual beam intensities of each isotope obtained from the database.
At the end of the implantation period (Fig.~\ref{fig:tb155} left), mainly due to the saturation and accumulation the short-lived isotopes $^{155}$Er,Ho,Dy, total activities of up to 10 GBq can be reached. 
Several days of cool-down are required to reduce the sample activity to manageable levels and for $^{155}$Tb activity to build up (Fig.~\ref{fig:tb155} right).

Provided that beam intensities are within the normal range, an implantation time of several hours is sufficient to accumulate 370 MBq (= $2.45\cdot 10^{14}$ atoms) $^{155}$Tb after a cool-down period of 5 days. The precise interval is calculated based on the yields obtained prior to the implantation.

\section{Conclusion}
The \textit{TRIUMF Isotope Database} is a tool for TRUMF's RIB facility users to plan and design experiments. It also provides resources for RIB and target development by compiling data on yield measurements, targets and simulated production rates. The database is constantly expanded with new yield results and updated simulation data sets.

\section*{Acknowledgements}
TRIUMF receives funding  through  the  National  Research  Council  of  Canada.   This work was also supported by the New Frontiers in Research Fund - Exploration  NFRFE-2019-00128 and the  Natural  Sciences  and  Engineering  Research  Council  of  Canada, (NSERC):  SAPIN-2021-00030.
Many thanks to the ISAC operations crew for their support.

\bibliographystyle{elsarticle-num}

\bibliography{emis2022_17}

\end{document}